\documentclass[journal]{IEEEtran}

\usepackage{cite}

\ifCLASSINFOpdf
   \usepackage[pdftex]{graphicx}
\else
\fi

\usepackage[cmex10]{amsmath}

\usepackage{array}
\usepackage{mdwmath}
\usepackage{mdwtab}

\usepackage{acronym}

\begin{document}
%
\title{Influence of Modeling Methods on the Estimation of the Nonlinear Noise Statistics Considering Joint PMD and Kerr Effects in Fiber Transmission Systems}
%
%
%

\author{Sina Fazel, Djalal-Falih Bendimerad, Nicola Rossi, Petros Ramantanis and Yann Frignac\thanks{S. Fazel was with Department of Physics and Electronics, CNRS SAMOVAR, T\'el\'ecom SudParis, Institut Polytechnique de Paris when this work was carried out. He is currently with Delft Center of Systems and Control (DCSC), Delft University of Technology, Delft, The Netherlands.}\thanks{D. Bendimerad was with Department of Physics and Electronics, CNRS SAMOVAR, T\'el\'ecom SudParis, Institut Polytechnique de Paris when this work was carried out. He is currently with The Optical Communication Technology Lab, Huawei Paris Research Centre, France}\thanks{N. Rossi was with Nokia Bell Labs when this work was carried out. He is currently with Alcatel Submarine Networks (ASN), 91620 Nozay, France.}\thanks{P. Ramantanis is with Nokia Bell Labs, 91620 Nozay, France.}\thanks{Y. Frignac is a Professor at the Department of Physics and Electronics, CNRS SAMOVAR, T\'el\'ecom SudParis, Institut Polytechnique de Paris, France. e-mail: yann.frignac@telecom-sudparis.eu}}

\acrodef{ASE}[ASE]{Amplified Spontaneous Emission}
\acrodef{AWGN}[AWGN]{Additive White Gaussian Noise}
\acrodef{BER}[BER]{Bit Error Rate}
\acrodef{CNLSE}[CNLSE]{Coupled NonLinear Schr\"odinger Equation}
\acrodef{DCF}[DCF]{Dispersion Compensating Fiber}
\acrodef{DGD}[DGD]{Differential Group Delay}
\acrodef{DM}[DM]{Dispersion Management}
\acrodef{DSP}[DSP]{Digital Signal Processing}
\acrodef{DU}[DU]{Dispersion Unmanaged}
\acrodef{FFT}[FFT]{Fast Fourier Transform}
\acrodef{FWM}[FWM]{Four Wave Mixing}
\acrodef{GN}[GN]{Gaussian Noise}
\acrodef{OOK}[OOK]{On Off Keying}
\acrodef{OSNR}[OSNR]{Optical Signal to Noise Ratio}
\acrodef{SMF}[SMF]{Single Mode Fiber}
\acrodef{PDM}[PDM]{Polarization Division Multiplexing}
\acrodef{PDM-QPSK}[PDM-QPSK]{Polarization Division Multiplexed Quadrature Phase Shift Keying}
\acrodef{QPSK}[QPSK]{Quadrature Phase Shift Keying}
\acrodef{PMD}[PMD]{Polarization Mode Dispersion}
\acrodef{QoT}[QoT]{Quality of Transmission}
\acrodef{RV}[RV]{Random Variable}
\acrodef{SDN}[SDN]{Software Defined Network}
\acrodef{SNR}[SNR]{Signal to Noise Ratio}
\acrodef{SSFM}[SSFM]{Split Step Fourier Method}
\acrodef{WDM}[WDM]{Wavelength Division Multiplexed}
\acrodef{PDF}[PDF]{Probability Density Function}
\acrodef{PRQS}[PRQS]{Pseudo-Random Quaternary Sequence}
\acrodef{CMA}[CMA]{Constant Modulus Algorithm}
\acrodef{XPI}[XPI]{Cross-Polarization Interference}
\acrodef{SCI}[SCI]{single Channel Interference}
\acrodef{XCI}[XCI]{intra polarization Cross Channel Interference}
\acrodef{SOP}[SOP]{State Of Polarization}
\acrodef{EDFA}[EDFA]{Erbium Doped Fiber Amplifier}
\acrodef{GVD}[GVD]{Group Velocity Dispersion}
\acrodef{MZM}[MZM]{Mach-Zehnder Modulator}
\acrodef{RRC}[RRC]{Root Raised Cosine}
\acrodef{LEAF}[LEAF]{Large Effective Area Fiber}
\acrodef{NLI}[NLI]{Non-Linear Interference}
\acrodef{NLIN}[NLIN]{Non-Linear Interference Noise}

\maketitle

\begin{abstract}
In the recent context of Software Defined Optical Network, the fast and accurate \ac{QoT} estimation of the transmission link is essential. Gaussian Noise models are shown to yield a fast estimation of the average \ac{QoT} derived from deterministic system parameters, but do not capture the \ac{QoT} variability. In order to assess numerically the stochastic joint effect of \ac{PMD} and Kerr nonlinearities, system designers generally use the \ac{SSFM} based on Manakov-\ac{PMD} equation neglecting the nonlinear-\ac{PMD} term which is faster than using \ac{CNLSE} and enough accurate for fiber with short birefringence correlation length (around less than 10m). In this work, we present insights of the way to tune the parameters of this Manakov-\ac{PMD} method and its limitation when seeking an accurate estimation of the \ac{NLI} noise statistical distribution for all fibers potentially installed in the current optical network. In particular we compare this Manakov-\ac{PMD} method results with respect to the one obtained by \ac{CNLSE} while varying the fiber birefringence correlation length, \ac{PMD} coefficient and the fiber type. Our results highlight a potential discrepancy of $0.5$ $dB$ in the estimation of the $Q^{2}$ factor in one span \ac{PDM-QPSK} transmission with optimal launch power per channel and yield guidelines to choose the most suitable numerical estimation method. 
\end{abstract}


%
\IEEEpeerreviewmaketitle

\section{Introduction}
%
%
%
%
\IEEEPARstart{T}{he} concept of \acp{SDN} has been recently proposed to provide the necessary flexibility and reliability to handle the increasing network traffic demand \cite{Nunes2014a}. In this context, a fast tool that could estimate the \ac{QoT} is of particular interest. The \ac{QoT} estimation is strongly motivated by network providers as a wise choice of the system margins that can ensure the network reliability with no additional cost brought by excessive opto-electronic regeneration transponders. In the fiber optic communications domain, recent work on such analytical or semi-analytical tools has been shown to provide promising solutions for that purpose, with the most prominent such model being the \ac{GN} model \cite{Poggiolini2014c}.

However, it has been recently shown that the \ac{BER} (or $Q^2$ factor) can be more accurately described by a random variable rather than a constant value due to the stochastic nature of the \ac{NLI}, which highlights variability in the \ac{GN} model performance estimation, particularly in \ac{DM} transmission systems \cite{Rossi2016a,Sina2017a}. One among several phenomena that can randomly influence the \ac{QoT} is \ac{PMD} \cite{Keiser2003c,Ramaswami2009b,Essiambre2010f}. Although recent advances in \ac{DSP} embedded in the coherent receivers can overcome the impairments induced by \ac{PMD} itself,  the interplay of \ac{PMD} and Kerr nonlinearity causes stochastic variations of transmission performance \cite{Menyuk2006a,Bertran-Pardo2010c,Wang1999a,Kudou1997b,Boroditsky2006b}. To assess \ac{PMD} and Kerr nonlinearity interaction, traditionally the well-known \ac{SSFM} has been used to numerically estimate the evolution of signals along the propagation based on the \ac{CNLSE}\cite{Weideman1986a,Agrawal2007a}. The \ac{SSFM} inherently requires a large amount of \ac{FFT} steps to fairly emulate both chromatic dispersion and Kerr nonlinearity. In such circumstances, \ac{PMD} adds more processing requirements due to an additional fiber splitting step and a massive set of random draws of fiber birefringence concatenations to properly investigate the whole probability distributions of the possible \ac{DGD} between signal polarization tributaries\cite{Serena2011c,Gao2012c}. Thus simulating transmission systems involving \ac{PMD} and Kerr nonlinearity implies a large amount of time-consuming computations. Manakov-\ac{PMD} method is offering an efficient way to reduce the computational complexity by emulating only the averaged Kerr nonlinear effects through the polarization states, assuming that the correlation length ($L_{corr.}$) of the fiber birefringence is much lower than the Kerr nonlinear length ($L_{Kerr}$)\cite{Marcuse1997a}. Authors in \cite{Marcuse1997a} have made the theoretical basis deriving from the \ac{CNLSE} an equivalent equation called Manakov-\ac{PMD} equation and have shown that its nonlinear-\ac{PMD} term can be fairly neglected by a numerical simulation in 1997 using legacy 5 Gbits/s NRZ single-channel and single-polarization transmission with computation constraints of the time imposing short sequence lengths and a few fiber random birefringence draws. In this paper we will refer to the manakov-\ac{PMD} method as the one describe in \cite{Marcuse1997a} while neglecting the nonlinear-\ac{PMD} term or equivalently the Eq. 68 of \cite{Menyuk2006a}. Later, many studies of joint \ac{PMD} and Kerr nonlinearity have been performed generally tackling very particular system configurations \cite{Serena2011c,Gao2012c}. While the computation of the signal propagation based on Manakov-\ac{PMD} equation neglecting the nonlinear-\ac{PMD} term is clearly less time consuming, it should be noticed this can not be blindly applied for all fiber types of the installed network as it may yields to \ac{NLI} noise estimation error.

In order to help system designers to find the most accurate and fast approach to estimate the \ac{QoT} variability, here we take profit of recent computational means and the above-mentioned \ac{SDN}-driven motivation to perform a massive amount of \ac{SSFM}-based \ac{WDM} transmission system simulations through both \ac{CNLSE} and Manakov-\ac{PMD} method with different fiber correlation lengths and the context of recent \acl{PDM} coherent transmission systems . Inspired by previous modeling efforts that separate the influence of the number of spans and other system parameters \cite{Bononi2012a, Vacondio2012b, Seve2013b} and in an effort to simplify our system under study, in this paper we assess the power-independent \ac{NLI} coefficient $a_{NL}$ as a \ac{RV}\cite{Rossi2016a} also depending on the fiber random birefringence concatenation  after a one-span transmission via either Manakov-\ac{PMD} or \ac{CNLSE} methods. The two methods are here specifically compared through the estimation of the average values and standard deviations of $a_{NL}$. Moreover, in our investigations the \ac{PMD} is taken into account only through its first order while varying the fiber \ac{PMD} coefficient and fiber types. 

The paper is organized as follows. In section 2 we review the numerical methods for the estimation of transmission quality in presence of \ac{PMD} and Kerr effect. This section contains \ac{SSFM} description accounting for \ac{PMD} and Kerr effect using either \ac{CNLSE} or Manakov-\ac{PMD} methods, fundamentals of \ac{GN} modeling to estimate \ac{NLI} noise accumulation, our simulation setup and finally \ac{NLI} noise statistics estimation process. Next in section 3, first we highlight the difference in \ac{NLI} noise statistics estimation using either \ac{CNLSE} or Manakov-\ac{PMD} equation while varying the number of birefringent plate employed to model the \ac{PMD} effect along the propagation link. Later, we analyze how the results are modified while changing the \ac{PMD} coefficient and fiber type. In the end, we give insights on choosing the most suitable method depending on the birefringence correlation length of the fiber to be modeled and quantify the estimation error made by using inadequate method.

\section{Transmission system performance estimation in presence of PMD and Kerr nonlinear effects}
\label{sec:sim_model}
\subsection{Split Step Fourier Method fiber propagation modeling in presence of PMD and Kerr nonlinearity}
Modeling the propagation of dual polarization modulated signals in the fiber generally relies on the simplified \ac{CNLSE} formalism, shown in Eq. (\ref{equ:cnlse}), which gives the evolution of the slow varying envelops $A_{x}(z,t)$ and $A_{y}(z,t)$ along the propagation distance $z$ assuming only loss, second order chromatic dispersion and Kerr nonlinear effects\cite{Agrawal2007a}. 
\begin{equation}
\resizebox{0.5\textwidth}{!}{$
\left\{\begin{matrix}
\frac{\partial A_x}{\partial z} +\beta_{1x} \frac{\partial A_x}{\partial t}+\frac{i \beta_{2x}}{2} \frac{\partial^2 A_x}{\partial t^2} + \frac{\alpha_x}{2} A_x=i \gamma (|A_x|^2+\frac{2}{3}|A_y|^2)A_x\\ 
\\
\frac{\partial A_y}{\partial z} +\beta_{1y} \frac{\partial A_y}{\partial t}+\frac{i \beta_{2y}}{2} \frac{\partial^2 A_y}{\partial t^2} + \frac{\alpha_y}{2} A_y=i \gamma(|A_y|^2+\frac{2}{3}|A_x|^2)A_y\\ 

\end{matrix}\right.
$}
\label{equ:cnlse}
\end{equation}
In the equation Eq. (\ref{equ:cnlse}), all $\beta_{n}(\omega)$ indicate the $n^{th}$ derivatives of the propagation constant $\beta$ as a function of $\omega$ around the overall signal center angular frequency $\omega_{0}$, $\alpha $ indicates the fiber attenuation coefficient and $\gamma$ is the nonlinear coefficient. Note that the $\beta_{2, x/y}$ and the $\alpha_{x/y}$ are here assumed equal for both polarization components (i.e. Polarization Dependent Loss is neglected). Note that $A_{x}(z,t)$ and $A_{y}(z,t)$ do not represent the envelops of a given channel but the total  \ac{WDM} field envelop of each polarization. Thus equation Eq. (\ref{equ:cnlse}) includes inherently \ac{FWM} for each polarization state.

The numerical calculation of the \ac{CNLSE} can be made by the use of \ac{SSFM} algorithm which successively applies nonlinear and dispersive steps \cite{Agrawal2007a,Jeruchim2006c}. Optimized \ac{SSFM} algorithms mainly consider nonlinear step that accumulates a constant amount of additional nonlinear phase $\phi_{NL}$ \cite{Sinkin2003b,Zhang2005a}. As the signal power is exponentially decreasing along the fiber, the emulated step length is thus increasing. It is worth noting that the accuracy of the \ac{SSFM} depends on the maximum nonlinear phase ($\phi_{NL,max}$) that can be accumulated in one step\cite{Agrawal2007a}.

On the other hand, we model the fiber under test as concatenation of birefringent  over the previous \ac{SSFM} fiber steps to model the \ac{PMD}. This results in a concatenation of successive plates each being with constant birefringence and randomly drawn birefringence axis. The length of fiber within which the birefringence properties can be considered constant is referred to as the correlation length $L_{corr.}$ \cite{Galtarossa2001a,Sikka1998a,Antonelli2008a,Galtarossa2000a,Galtarossa2000b,Galtarossa2000c,Galtarossa2000d}. For accurate estimation of the \ac{PMD}, length of the plates has to be chosen equal to the $L_{corr.}$ of a real fiber. For instance, some previous measurements indicate (see Tab. \ref{tab:lcorr}) a $L_{corr.}$ from about 10 to 100 meters respectively corresponding to 10 000 to 1000 plates in order to emulate a 100km-long fiber span. For a reasonable time consumption, \ac{SSFM} simulations based on the \ac{CNLSE} are generally using a lower number of plates yet with the cost of a loss of accuracy in the calculation (see section \ref{nplates_variation}).
\begin{table}[]
	\begin{center}
		\begin{tabular}{|l|l|l|}
			\hline
			Fiber type    & $L_{corr.}$[m]              &   \ac{PMD} coefficient $[ps/\sqrt{km}]$           \\ \hline
			G.652     &  $23$  & 0.05 \\ \hline
			G.653 &  10 & 0.06\\ \hline
			G.655 &  $19$ & 0.05\\ \hline
		\end{tabular}
		\caption{Examples of correlation length $L_{corr.}$ and \ac{PMD} coefficient measurements from literature \cite{Galtarossa2000b,Galtarossa2000d}.}
		\label{tab:lcorr}
	\end{center}
\end{table}

While the correlation length is much lower than the nonlinear Kerr length ($L_{corr.} \ll L_{Kerr}$), the nonlinear effects on signals can be modeled as somehow averaged over polarizations, following the traditional Manakov-\ac{PMD} equation (see Eq. \ref{equ:manakov}) instead of the \ac{CNLSE} (derived from Eq. 68 in \cite{Menyuk2006a}). We recall as previously explained in the introduction that the nonlinear-\ac{PMD} term is neglected as it is described in \cite{Menyuk2006a}. This method has the advantage of effectively reducing the simulation computation time \cite{Marcuse1997a,Wai1996a,Wai1997a}.
\begin{equation}
\resizebox{0.5\textwidth}{!}{$
\left\{\begin{matrix}
\frac{\partial A_x}{\partial z} +\beta_{1x} \frac{\partial A_x}{\partial t}+\frac{i \beta_{2x}}{2} \frac{\partial^2 A_x}{\partial t^2} + \frac{\alpha_x}{2} A_x=i \gamma \frac{8}{9}(|A_x|^2+|A_y|^2)A_x\\ 
\\
\frac{\partial A_y}{\partial z} +\beta_{1y} \frac{\partial A_y}{\partial t}+\frac{i \beta_{2y}}{2} \frac{\partial^2 A_y}{\partial t^2} + \frac{\alpha_y}{2} A_y=i \gamma \frac{8}{9}(|A_y|^2+|A_x|^2)A_y\\ 

\end{matrix}\right.
$}
\label{equ:manakov}
\end{equation}

As the terms $(|A_x|^2)A_x$ and $(|A_y|^2)A_y$ in the \ac{CNLSE} equation (Eq. (\ref{equ:cnlse})) are related to  Kerr nonlinearities within each polarization tributary while the terms $(\frac{2}{3}|A_y|^2)A_x$ and $(\frac{2}{3}|A_x|^2)A_y$ are referring to Cross-Polarization nonlinearities, we can observe that the $\frac{8}{9}$ coefficient in the Manakov-\ac{PMD} equation (Eq. (\ref{equ:manakov})) undermines the weights of intra-tributary nonlinearities of \acs{PDM} tributaries with respect to the weights of cross polarization nonlinear effects.

\subsection{Introducting $Q^{2}$ factor and $a_{NL}$, nonlinear noise coefficient in Gaussian noise modeling}

\label{sec:gn}

The \ac{BER} of systems impacted by \ac{AWGN} has been studied extensively in classical digital communications and simple closed form formulas exist for the \ac{BER} as a function of the \ac{SNR} for various modulation formats. E.g. for \ac{QPSK} modulation, the \ac{BER} is given by the well-known formula \cite{Proakis2000}

\begin{equation}
BER = \frac{1}{2}\operatorname{erfc} \left( {\sqrt {\frac{{SNR}}{2}} } \right)
\label{BERvsSNRforQPSK}
\end{equation}

where the \ac{SNR} is defined as

\begin{equation}
SNR = \frac{{{P}}}{{{P_N}}}
\label{SNRDef}
\end{equation}

with ${{P}}$ the signal power and ${{P_N}}$ the total noise power or noise variance.

In optical 10Gbps \ac{OOK} systems, supposing Gaussian statistics for the two symbols, the \ac{BER} could be given as a function of the Q factor using the formula \cite{Agrawal2005}

\begin{equation}
BER = \frac{1}{2}\operatorname{erfc} \left( {\frac{Q}{{\sqrt 2 }}} \right)
\label{QFactorDef}
\end{equation}

Even though 10Gbps \ac{OOK} systems have almost entirely given their place to coherent 100Gbps systems, in optical communication systems it is customary to convert \ac{BER} (e.g. obtained by lab measurements by Monte-Carlo error counting) into an equivalent $Q_{BER}$ (i.e. $Q$ factor obtained from \ac{BER}) by inverting eq. \ref{QFactorDef}

\begin{equation}
{Q_{BER}} = \sqrt 2 {\operatorname{erfc} ^{ - 1}}\left( {2BER} \right)
\label{Q2BERDef}
\end{equation}

As it can be seen from Eqs. \ref{BERvsSNRforQPSK} and \ref{Q2BERDef}, for an optical coherent system with \ac{QPSK} modulation and Gaussian noise statistics, we expect $Q_{BER}^2 = SNR$.

As shown in \cite{Carena2010b}, \cite{Grellier2011}, \cite{Bononi2012a}, the distortion due to the nonlinear Kerr effect in coherent \ac{DU} systems can be treated as an additive \ac{AWGN} term, i.e. assuming no correlation between \ac{NLI} "noise" and \ac{ASE} noise, the \ac{SNR} at the receiver can be written as\cite{Vacondio2012b}

\begin{equation}
SNR = \frac{{{P_s}}}{{{P_{ASE}} + {P_{NL}} + {P_{TRX}}}}
\label{SNRGN}
\end{equation}

where ${{P}}$ is the received signal power, ${{P_{ASE}}}$ is the noise variance added by the amplifiers, ${{P_{TRX}}}$ accounts for the distortion due to transponder imperfections and ${{P_{NL}}}$ is the \ac{NLI} "noise" variance. In the same (\ac{GN} model) context, the \ac{NLI} "noise" variance has been shown to scale cubically with power, i.e.

\begin{equation}
{P_{NL}} = {a_{NL}} {P^3}
\label{Pnl}
\end{equation}
where ${a_{NL}}$ is a power independent \ac{NLI} coefficient.
By using Eq. \ref{Pnl} in \ref{SNRGN} and taking the inverse, for a N-span heterogeneous system\cite{Seve2013b} we get

\begin{equation}
\frac{1}{{SNR}} = \frac{1}{{SN{R_{TRX}}}} + \sum\limits_{k = 1}^N \left ( {\frac{{N{F_k} \cdot h\nu \left( {{G_k} - 1} \right)B}}{{P_k}}} + {a_{{NL}_k}P_k^2} \right )
\label{InverseSNRGN}
\end{equation}

where $P_k$ is the power at the input of span k, $NF_k$ and $G_k$ are the kth amplifier noise figure and gain, $h$ is the Planck constant, $\nu$ is the light frequency and $B$ is the signal bandwidth (e.g. equal to the symbol-rate for Nyquist signals). From Eq. \ref{InverseSNRGN} we note that the quantities accumulating along the line are the inverse \acp{SNR}, due to either \ac{ASE}, nonlinearities or transponder imperfections.

For some common assumptions, i.e. a) ideal transponders (${P_{TRX}}=0$ and $SNR_{TRX}= + \infty $), b) identical spans and injection powers c) identical amplifiers (i.e. same noise figures) with gains compensating exactly for the signal span loss d) a supra-linear accumulation of the \ac{NLI} noise along the line, Eqs. \ref{SNRGN}, and \ref{InverseSNRGN} yield \cite{Bononi2012a}

\begin{equation}
SNR = \frac{P}{{N \cdot NF \cdot h\nu \left( {G - 1} \right)B + {\alpha _{NL}}{N^{1 + \varepsilon }}{P^3}}}
\label{SNRGNAssumptions1}
\end{equation}
\begin{figure*}[t]
	\centering
	\begin{tabular}{@{}c@{}}
		\includegraphics[width=.7\linewidth]{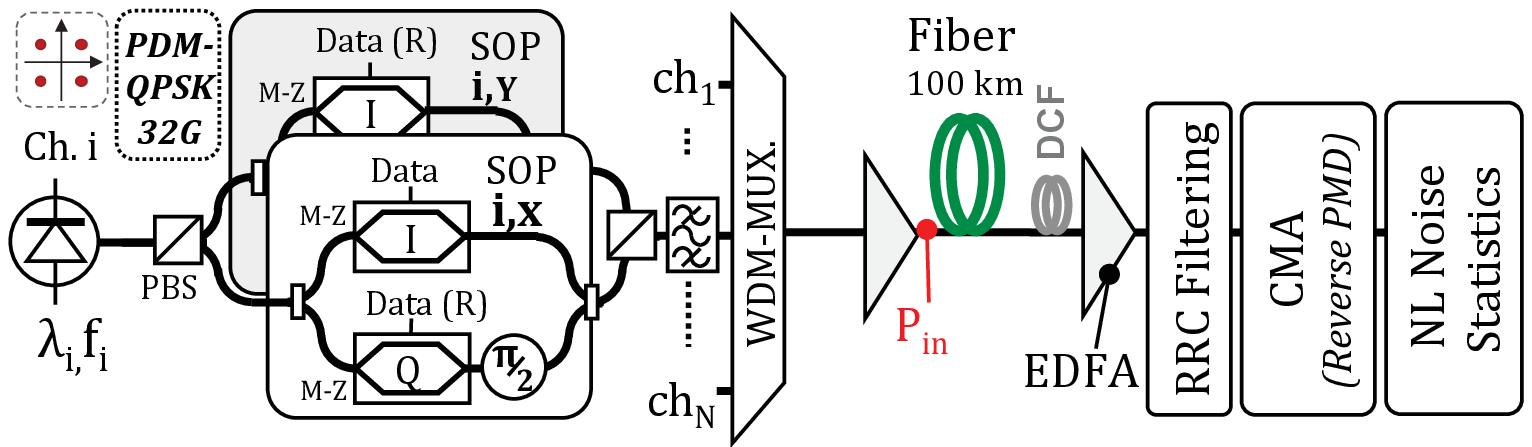} 
		\includegraphics[width=.2\linewidth]{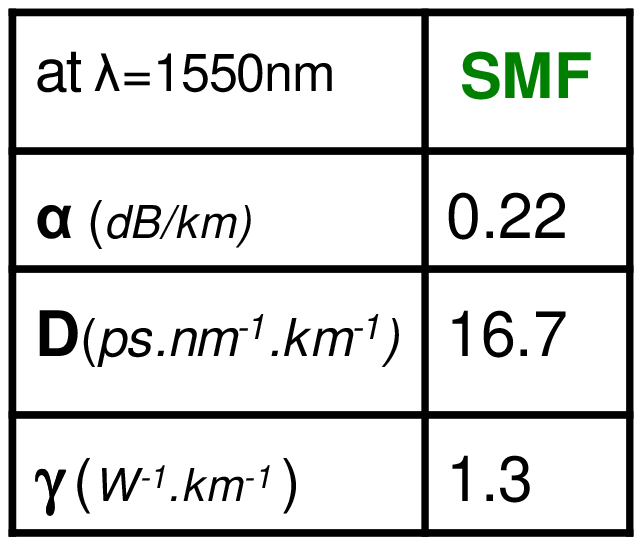} \\
		\hspace{3cm}\small (a)  \hspace{9cm} \small (b) 
	\end{tabular}
	\caption{Transmission system simulation setup (\textit{PBS : Polarization Beam Splitter, M-Z : Mach-Zehnder, SOP : State Of Polarization, Mux. : multiplexer, EDFA : Erbium Doped Fiber Amplifier)}.}
	\label{fig:setup}
\end{figure*}
where $\alpha_{NL}$ is a system-dependent constant that depends on the span characteristics and $\varepsilon$ takes into account the \ac{NLI} noise correlation between spans\footnote{Typical values can be found in \cite{Bononi2012a} (Fig. 7) ${\alpha _{NL}} = 3.95 \cdot {10^{ - 4}}m{W^{ - 2}}$ and $\varepsilon  = 0.22$}. $ {a _{NL}} $ can be then retrieved as $  {a _{NL}} = {\alpha _{NL}}{N^{1 + \varepsilon }}$.

As an example, we consider a system with $N$ spans, operating at the optimal power $P_{NLT}$. The \ac{SNR} is given by \cite{Bononi2012a} (Eq. 14)

\begin{equation}
SN{R_{opt}} = \frac{{{2^{\frac{2}{3}}}}}{{3 \cdot {\alpha _{NL}}^{\frac{1}{3}}{N^{1 + \frac{\varepsilon }{3}}}{{\left[ {NF \cdot hv \cdot \left( {G - 1} \right)B} \right]}^{\frac{2}{3}}}}}
\label{SNRGNAssumptions}
\end{equation}

As also mentioned in \cite{Poggiolini2014c}, eq. \ref{SNRGNAssumptions} states that the optimal \ac{SNR} is proportional to ${{\alpha _{NL}}^{-\frac{1}{3}}}$ or proportional to $ - \frac{1}{3}{\alpha _{NL}}$ in a dB scale. This means e.g. that a $3$ $dB$ increase in ${\alpha _{NL}}$ would have an impact of $1$ $dB$ decrease on the \ac{SNR} or directly on the $Q_{BER}^2$ for \ac{QPSK} modulation.

In this paper, we will not discuss on the validity of the \ac{GN} model however we focus on the impact of joint stochastic \ac{PMD} and Kerr effect on the variation of $a_{NL}$ coefficient which is then practical to quickly estimate the transmission \ac{QoT} and its variability.

\subsection{Our simulation setup}

In this work, we perform numerical simulations using Central and Graphics Processing Unit (CPU/GPU) computations to emulate the transmission system setup shown in Fig. \ref{fig:setup}. Our transmitter consists in a classical \ac{WDM} \ac{PDM-QPSK} optical transmitter multiplexing 21 channels. Each \ac{WDM} channel is generated by multiplexing signals on two orthogonal polarizations that are modulated separately by In-Phase (I) and Quadrature (Q) \ac{QPSK} \acp{MZM} with modulation rate $R=32$ GBauds. One PDM tributary results in an optical signal carrying 16384-long ($4^7$) DeBruijn sequence generated from a \ac{PRQS} of symbols oversampled 128 times. All \ac{WDM} channels and polarization tributaries are based on different \ac{PRQS} streams. The different \ac{WDM} laser signals are multiplexed following a 50 GHz frequency spacing and their Stokes's vectors are randomly drawn on the Poincar\'e Sphere leading to random \acp{SOP} for each channel. We use \ac{RRC} pulse shaping with a roll-off factor of 0.1. 

As discussed in the previous section, we only consider a single-span transmission of 100 km fiber with launched power per channel being $P_{in}=0$ $dBm$ (since $a_{NL}$ is power independent). We also assume ideal flat gain and noiseless black-box amplifiers to model \acp{EDFA}. As described in section \ref{sec:sim_model}, the numerical simulation of the signal propagation along the fiber is made following either \ac{CNLSE} or Manakov-\ac{PMD} methods considering only loss, \ac{GVD}, \ac{PMD} and Kerr nonlinear effects (based on open-source Optilux simulation software \cite{Optilux2009a}). Here we consider \ac{SMF} with nonlinear refractive index coefficient $n_2 = 2.6\times10^{-20}$ $m^2/W$  and effective area $A_{eff}=80 \times 10^{-12}$ $m^2$ (nonlinear coefficient $\gamma = 1.3$ $W^{-1}\cdot km^{-1}$) and with \ac{GVD} coefficient $D=16.7$ $ps.nm^{-1}\cdot km^{-1}$. The \ac{SSFM} step length is calculated in such a way that a maximum nonlinear phase ($\Delta \phi_{max}$) of $5 \cdot 10^{-4}$ rad is accumulated at each step. The fiber \ac{PMD} modeling is performed by a concatenation of $N_{p}$ birefringent plate with \ac{DGD} according to the fiber \ac{PMD} coefficient. 

At the end of the link, in order to directly estimate the \ac{NLI} noise distortions, we numerically apply a full dispersion compensation (equivalent to an ideal \ac{DCF} as shown in Fig. \ref{fig:setup}), matched \ac{RRC} filtering, ideal (data-aided) phase and polarization recovery to the signals. Here, the polarization recovery is artificially made in simulation by applying the linear reverse matrix of the whole plates concatenation (called "Reverse-PMD" block in Fig. \ref{fig:setup}). In this way the joint PMD-nonlinear distortions on signals are investigated without being altered  by the coherent mixer or the equalization algorithms at the receiver side. The $a_{NL}$ coefficient estimation is then performed on the two demultiplexed polarization tributaries of the central channel as described in the following section \ref{sec:anl_meas}.

\subsection{NL noise statistic assessment parameters}

\label{sec:anl_meas}

\begin{figure}
	\centering
	\includegraphics[width=.8\linewidth]{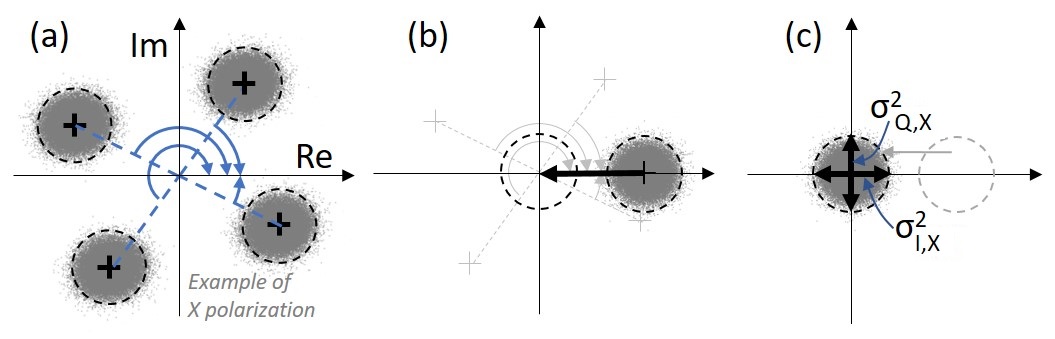}
	\caption{$a_{NL}$ estimation principle steps (for the central channel). (a) step 1 : finding the average complex number for each \ac{QPSK} state and canceling the average state phase, (b) step 2 : canceling the real part of average superimposed \ac{QPSK} states, (c) step3:  I,Q-variances estimation needed for $a_{NL}$ calculation.}
	\label{fig:anl}
\end{figure}

The estimation of the $a_{NL}$ coefficient is illustrated in Fig. \ref{fig:anl}. The $a_{NL}$ coefficient is calculated through $a_{NL}=(\sigma_{I,X}^{2}+\sigma_{Q,X}^{2}+\sigma_{I,Y}^{2}+\sigma_{Q,Y}^{2})/<P^{3}>$ where $P$ is the channel power (including both polarizations), and $\sigma^{2}$ stands for the variance of $T$-spaced sampled numerical signal ($T=1/R$). $\sigma_{I,X or Y}^{2}$ and $\sigma_{Q,X or Y}^{2}$ are estimated following the principle shown in  Fig. \ref{fig:anl}. Complex samples of each received \ac{QPSK} state are rotated by canceling the phase and then the modulus of the state average complex number (see Fig. \ref{fig:anl}a). Then we estimate the \ac{PDF} of the gathered $16384$ samples of all states (see Fig. \ref{fig:anl}c) and derive the real and imaginary variances which are  equivalent to the In-Phase and Quadrature variances of the received \ac{QPSK} states (see Fig. \ref{fig:anl}b). 

\begin{figure*}[t]
	\centering
	\begin{tabular}{@{}c@{}}
		\includegraphics[width=.8\linewidth]{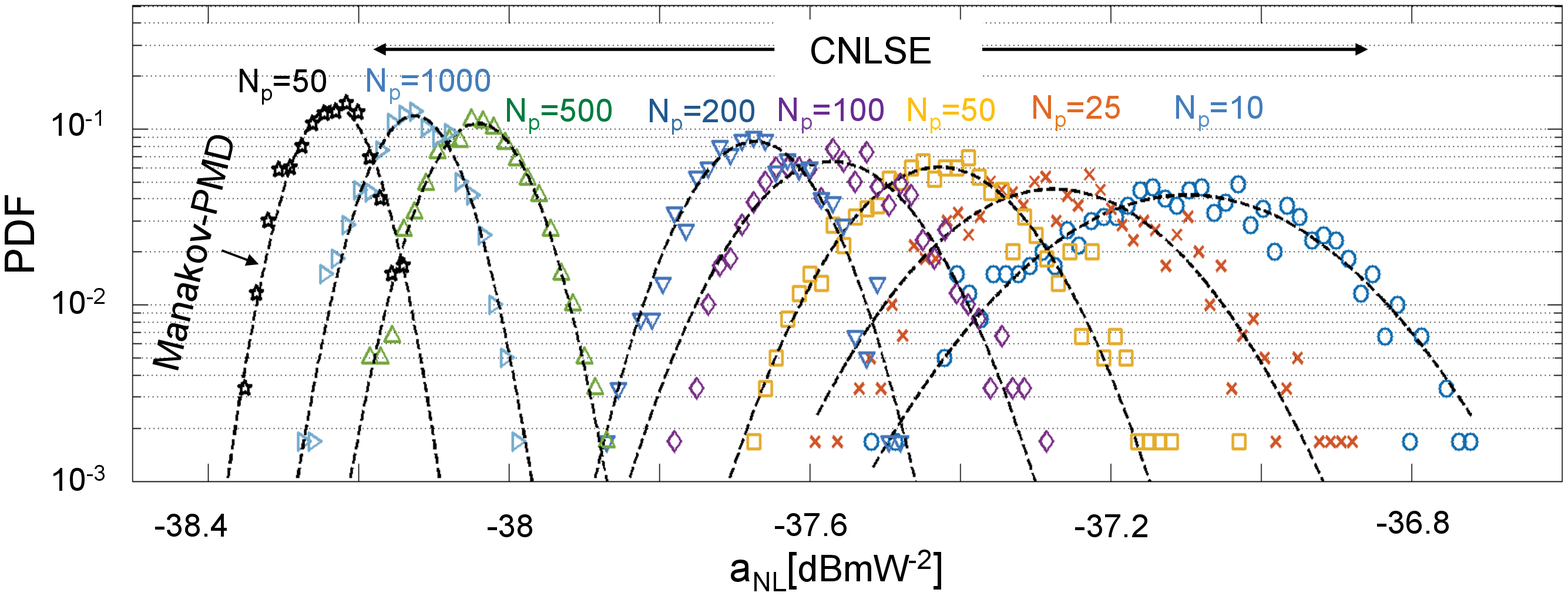}\\
		\hspace{1cm}\small (a) \\
		\includegraphics[width=.48\linewidth]{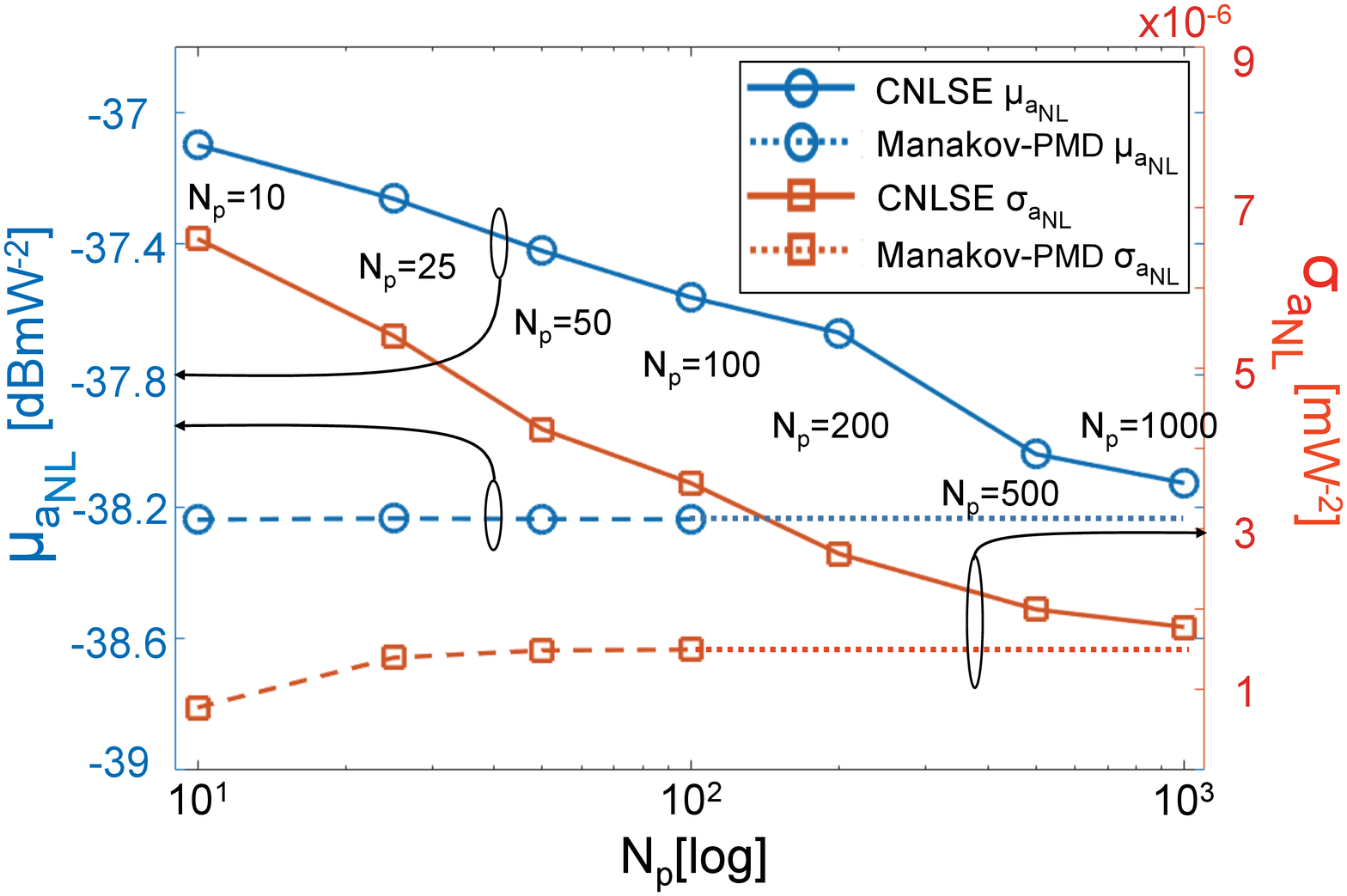}
		\includegraphics[width=.52\linewidth]{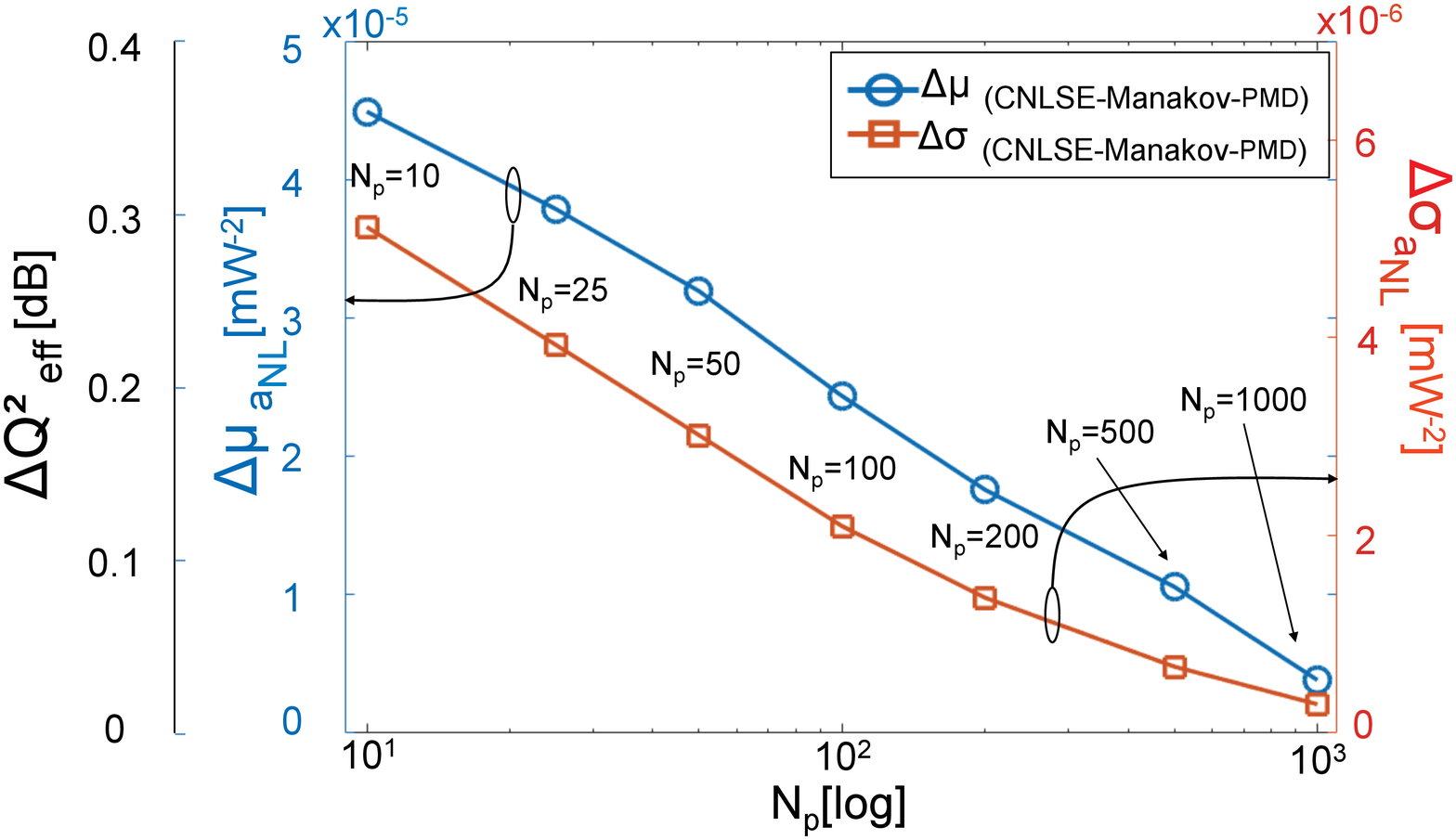}\\
		\hspace{0.2cm} \small (b)  \hspace{9.2cm} \small (c) 
	\end{tabular}
	\caption{(a)  \ac{PDF} of $a_{NL}$ over a set of 600 transmissions with \ac{PMD} coefficient of 0.13 $ps/\sqrt{km}$. (b) Boundary estimation based on $N_p$. (c) Estimation difference between \ac{CNLSE} and Manakov-PMD methods.}
	\label{fig:mancnlse_smf}
\end{figure*}
In order to reach our objectives for estimation of the transmission performance variability caused by both \ac{PMD} and Kerr effects using \ac{CNLSE} and Manakov-\ac{PMD}, we then estimate $a_{NL}$ over a set of 600 transmissions of 100km-long fiber span, each emulated with a random draw of $N_{p}$ birefringent plates. This number of draws is chosen as a trade-off between a fair estimation of the Maxwellian \ac{DGD} probability density function and a reasonable computation time. Finally from the \ac{PDF} of this set of 600 obtained $a_{NL}$ values, we extract the average value of $a_{NL}$ (from values in $mW^{-2}$) that we will refer to as $\mu _{a_{NL}}$ (converted to dB-scale and counted in $dBmW^{-2}$ following $10 \cdot log_{10} [mW^{-2}]$) and its standard deviation noted as $\sigma _{a_{NL}}$ in the following.

\section{Comparison of CNLSE and Manakov-PMD methods to estimate nonlinear noise statistics}

\subsection{CNLSE vs. Manakov-PMD estimation using one-span SMF transmission}
\label{nplates_variation}

Fig. \ref{fig:mancnlse_smf}a shows the \ac{PDF} of $a_{NL}$ over a set of 600 transmissions of 100km-long \ac{SMF} span using Manakov-\ac{PMD} equation with $N_{p}=50$ and \ac{CNLSE} with $N_{p}=$10, 25, 50, 100, 200, 500 and 1000  (see section \ref{sec:sim_model} for method descriptions). Here, we consider a \ac{PMD} coefficient of 0.13 $ps/\sqrt{km}$. For all $a_{NL}$ probability distributions markers refer to the histogram obtained from \ac{SSFM} simulation results (with different $N_{p}$) while dashed curves correspond to associated Gaussian fittings.

We can observe that when the number of plates is increasing, the  $\mu _{a_{NL}}$ and $\sigma _{a_{NL}}$ values obtained using \ac{CNLSE} tend to decrease towards the ones obtained using Manakov-\ac{PMD} method. This observation as already been shown and explained by \cite{Marcuse1997a} due to the fact that the impact of the nonlinear-\ac{PMD} term vanishes when the birefringence correlation length tends to 0 (or equivalently $N_{p}$ tends to infinite). Note that for the results using Manakov-\ac{PMD} method (we recall without nonlinear-\ac{PMD} term), we still observe a residual variability coming from the impact of linear stochastic \ac{PMD} on Kerr nonlinear distortions.
Besides our results quantify the difference of nonlinear noise statistics obtained using the two methods and depending on the actual birefringence correlation length of the considered fiber.
\begin{figure*}[h!]
	\centering
	\begin{tabular}{@{}c@{}}
		\includegraphics[width=.5\linewidth]{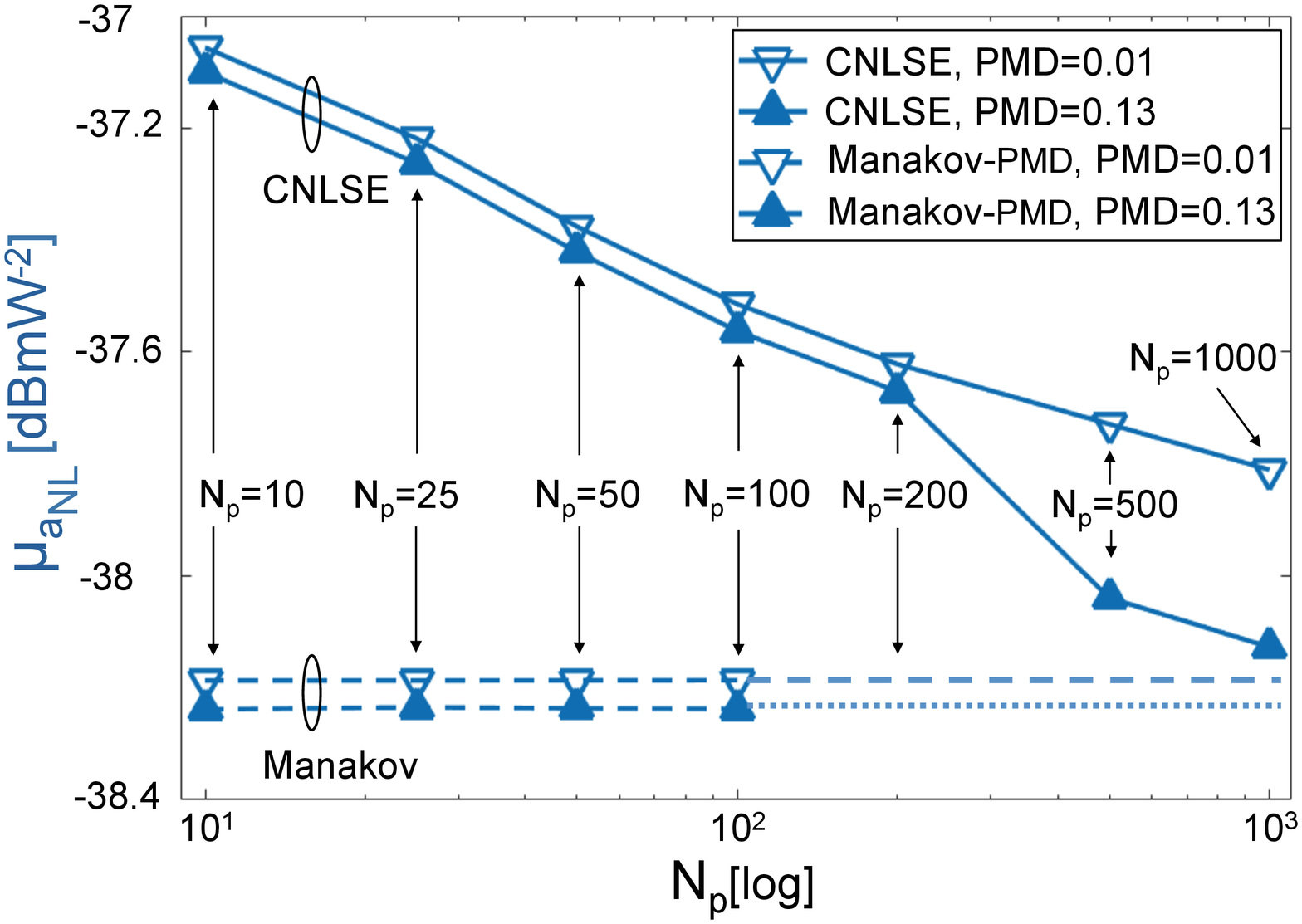}
		\includegraphics[width=.51\linewidth]{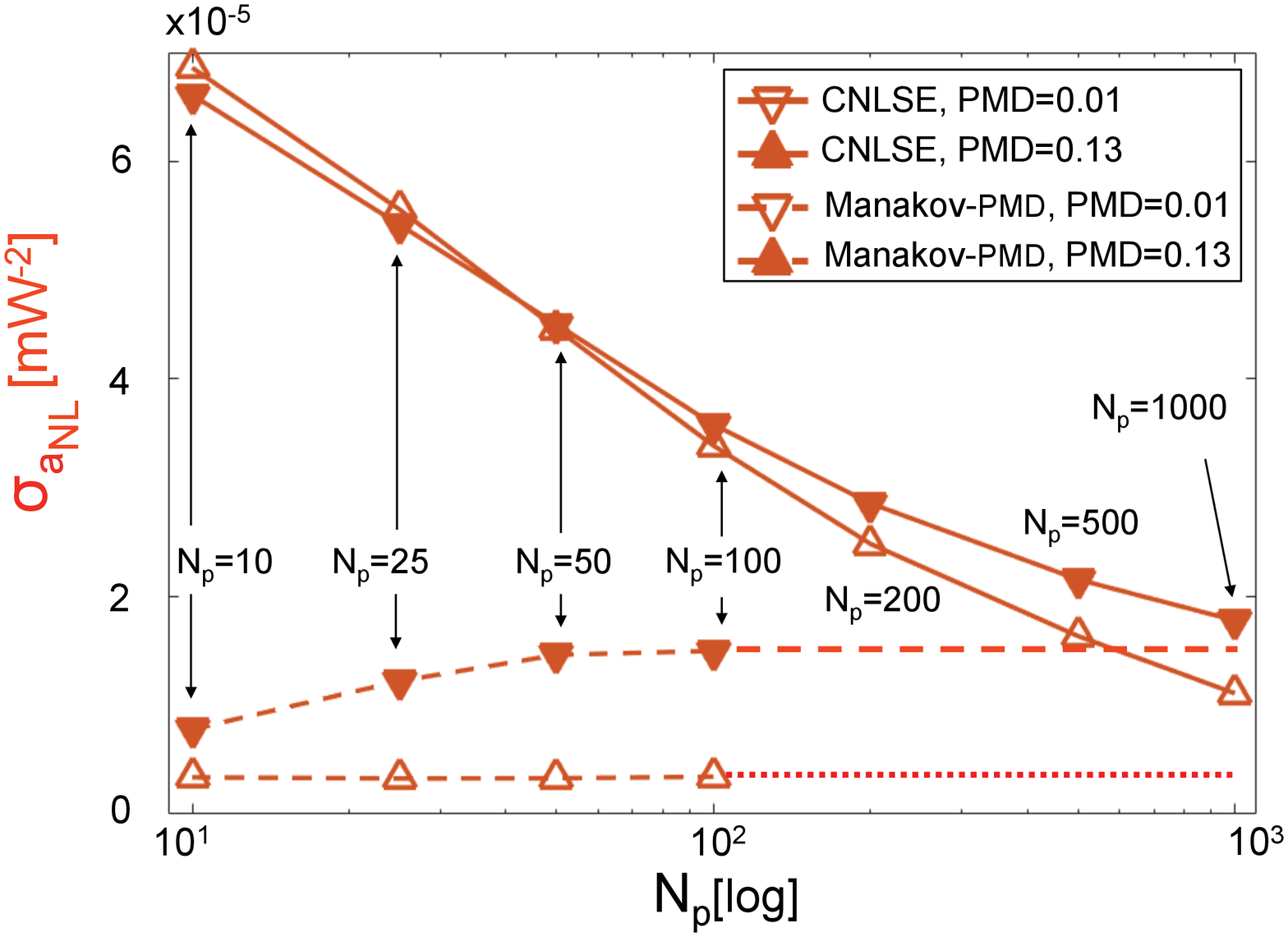}
		\\
		\hspace{1cm}\small (a) \hspace{9cm} \small (b)  
	\end{tabular}
	
	\caption{$\mu _{a_{NL}}$ (a) and $\sigma _{a_{NL}}$ (b) vs. $N_{p}$ using CNSLE and Manakov-PMD methods for SMF with \ac{PMD} coefficients of 0.01 and 0.13 $ps/\sqrt{km}$.}
	\label{fig:mancnlse_pmd}
\end{figure*}

Manakov-\ac{PMD} method with $N_{p}=50$ exhibits $\mu _{a_{NL}}=-38.2$ $dBmW^{-2}$ and $\sigma _{a_{NL}}=1.5 \cdot 10 ^{-6}$ $mW^{-2}$ while, for instance, \ac{CNLSE} with $N_{p}=$ 10 and 1000 exhibits $\mu _{a_{NL}}=-37.1$ $dBmW^{-2}$, $\sigma _{a_{NL}}=6.6 \cdot 10 ^{-6}$ $mW^{-2}$ and $\mu _{a_{NL}}=-38.1$ $dBmW^{-2}$, $\sigma _{a_{NL}}=1.8 \cdot 10 ^{-6}$ $mW^{-2}$ respectively. The result show that in \ac{CNLSE} simulations, the total variation of $\mu _{a_{NL}}$ is about $1.6$ $dB$ considering different $N_p$, while Manakov-\ac{PMD} provides almost constant results changing $N_p$.
\begin{figure*}[t]
	\centering
	\begin{tabular}{@{}c@{}}
		\includegraphics[width=.5\linewidth]{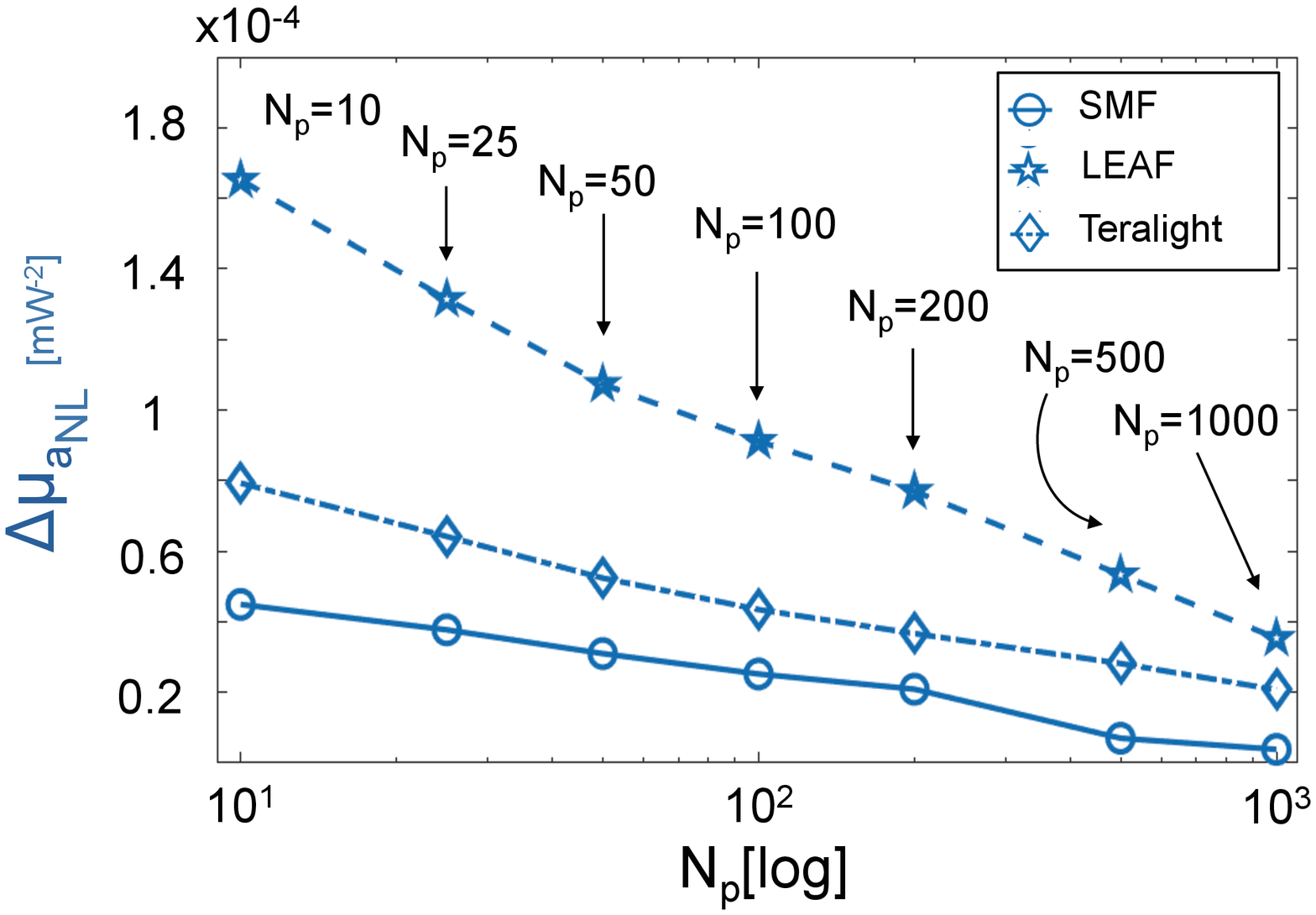}
		\includegraphics[width=.5\linewidth]{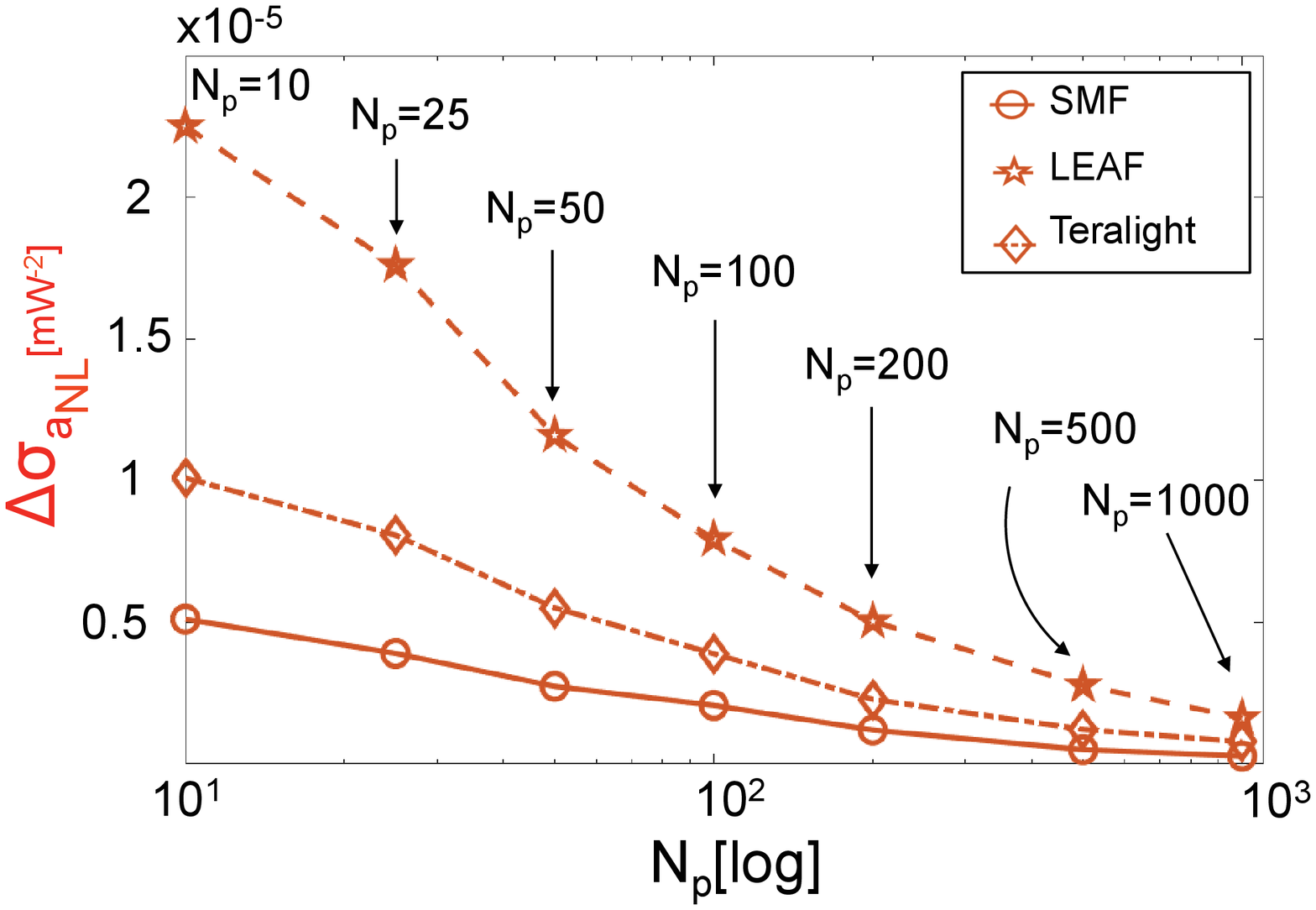}
		\\
		\hspace{1cm}\small (a) \hspace{9cm} \small (b)
	\end{tabular}
	
	\caption{$\Delta \mu _{a_{NL}}$ (a) and $\Delta \sigma _{a_{NL}}$ (b) vs. $N_{p}$ using SMF, LEAF and Teralight fiber.}
	\label{fig:mancnlse_fibers}
\end{figure*}

This observation is more accurately quantified in Fig. \ref{fig:mancnlse_smf}b where we report the values of $\mu _{a_{NL}}$ (left Y-axis) and $\sigma _{a_{NL}}$ (right Y-axis) as a function of $N_{p}$ in logarithmic scale for Manakov-\ac{PMD} method in dashed lines and \ac{CNLSE} method in solid lines. Here we can observe that Manakov-\ac{PMD} method provides constant $\mu _{a_{NL}}$ as a function of the number of plates varying from $N_{p}=10$ to $100$. This observation is somehow intuitive as Manakov-\ac{PMD} is based on averaging Kerr nonlinearity over polarization states. Then we extrapolate the same $\mu _{a_{NL}}$ value for higher number of plates always using Manakov-\ac{PMD} method (indicated by a dotted dashed line). Besides, considering Manakov-\ac{PMD} method, we equally note that the $\sigma _{a_{NL}}$ remains constant for $N_{p}$ values higher than around 25 or 50 plates. Therefore we can assume that using at least $N_{p}=50$ is enough to correctly estimate $\mu _{a_{NL}}$ and $\sigma _{a_{NL}}$ in the context of Manakov-\ac{PMD} modeling. For the \ac{CNLSE} method, we quantify more accurately how the $\mu _{a_{NL}}$ and $\sigma _{a_{NL}}$ values from Fig. \ref{fig:mancnlse_smf}a are decreasing toward the values obtained through Manakov-\ac{PMD} method when $N_{p}$ is increasing by plotting $\Delta \mu _{a_{NL}}=\mu _{a_{NL_{CNLSE}}} - \mu _{a_{NL_{Manakov}}}$  and $\Delta \sigma _{a_{NL}}=\sigma _{a_{NL_{CNLSE}}} - \sigma _{a_{NL_{Manakov}}}$ on Fig. \ref{fig:mancnlse_smf}c. From this results we can discuss on two typical situations. For example, to model an hypothetical fiber having a $L_{corr.}=2$ km, \ac{CNLSE} method with $N_{p}=50$ will provide more accurate results than Manakov-\ac{PMD} method. However by using Manakov-\ac{PMD} method $\mu _{a_{NL}}$ and $\sigma _{a_{NL}}$ will be underestimated  by $0.8$ $dB$ and $2.75 \cdot 10^{-6}$ $mW^{-2}$ respectively, thus yielding optimistic estimations. On the other hand for a fiber having $L_{corr.}=100$ m or less, it is better to use Manakov-\ac{PMD} method with only $N_{p}=50$ instead of \ac{CNLSE}  with $N_{p}=1000$ as Manakov-\ac{PMD} method provides similar results in largely less simulation time. Finally in order to rapidly estimate the $\Delta \mu _{a_{NL}}$ and $\Delta \sigma _{a_{NL}}$ for any $N_{p}$, we propose an empirical law for \ac{SMF} by driving and exponential fitting over the results of Fig. \ref{fig:mancnlse_smf}c following $\Delta \mu _{a_{NL}(N_p)} =  10^{-5} \cdot \exp(-10^{-2}  \times N_p) $ and $\Delta \sigma_{a_{NL}(N_p)} = 3.8  \cdot 10^{-6}  \cdot \exp(-2 \cdot 10^{-2}  \times N_p)$. 

As a preliminary conclusion to this part, one can note that in order to model fibers with a lower \ac{PMD} coefficient and shorter correlation length, Manakov-\ac{PMD} method is relevant as being enough accurate and gaining simulation time while attempting to model fibers with a higher \ac{PMD} coefficient may imply the use of \ac{CNLSE} method depending on the actual value of fiber $L_{corr.}$.

\subsection{Influence of fiber PMD coefficient}
\label{pmd_variation}
In this section we investigate the impact of the fiber \ac{PMD} coefficients on the previously estimated $\mu _{a_{NL}}$ and $\sigma _{a_{NL}}$ using both \ac{CNLSE} and Manakov-\ac{PMD} methods.
In Fig. \ref{fig:mancnlse_pmd}a and Fig. \ref{fig:mancnlse_pmd}b, we show the $\mu _{a_{NL}}$ and $\sigma _{a_{NL}}$ respectively as a function of $N_{p}$ for both methods with two different \ac{PMD} coefficients equal to 0.01 $ps/\sqrt{km}$ (empty downward triangles) and 0.13 $ps/\sqrt{km}$ (full upward triangles).

As the Fig. \ref{fig:mancnlse_pmd}a shows, by increasing the $N_p$, the convergence of $\mu _{a_{NL}}$ obtained with \ac{CNLSE} method towards the value achieved with Manakov-\ac{PMD} is approximately at $N_p=1000$ for \ac{PMD}=0.13 $ps/\sqrt{km}$ (full upward triangles), while we need higher $N_p$ values for \ac{PMD}=0.01 $ps/\sqrt{km}$ (empty downward triangles). This is due to the fact that having higher \ac{PMD} coefficient implies more efficient nonlinear averaging  in \ac{CNLSE} method and thus converge faster toward Manakov-\ac{PMD} results when increasing $N_p$. 

Moreover, as fiber \ac{DGD} follows a Maxwellian distribution \cite{Yan2004b}, higher \ac{PMD} values provide larger \ac{DGD} distribution thus the system variability due to \ac{PMD} is expected to increase. This is apparent through Fig. \ref{fig:mancnlse_pmd}b, where $\sigma _{a_{NL}}$ for \ac{PMD}=0.13 $(ps/\sqrt{km})$(full upward triangles) is higher than $\sigma _{a_{NL}}$ in \ac{PMD}=0.01 $(ps/\sqrt{km})$(empty downward triangles). The $\sigma _{a_{NL}}$ difference between two \ac{PMD} values is not clearly visible for $N_p=100$ considering \ac{CNLSE} method. However by using Manakov-\ac{PMD} method, the system variability can be more accurately evaluated for $N_p>50$. 

When decreasing the \ac{PMD} coefficients the evolution of $\mu _{a_{NL}}$ and $\sigma _{a_{NL}}$ with $N_{p}$ follows the same trends as the ones of Fig. \ref{fig:mancnlse_smf}b however we can note that the results obtained using \ac{CNLSE} methods converge toward the results obtained by Manakov-\ac{PMD} for larger $N_{p}$. 
Moreover, as the Fig. \ref{fig:mancnlse_pmd}b shows, the difference between $\sigma _{a_{NL}}$ values obtained by Manakov-\ac{PMD} or \ac{CNLSE} methods is lower for higher fiber \ac{PMD} coefficient values. However, when emulating a fiber with higher \ac{PMD} coefficient,  $\Delta \sigma _{a_{NL}}$ ($\sigma _{a_{NL_{CNLSE}}} - \sigma _{a_{NL_{Manakov}}}$) decreases. 

\subsection{Influence of fiber type used}
\label{fiber_variation}

In this section we investigate the difference between \ac{CNLSE} and Manakov-\ac{PMD} estimations by varying the fiber type.  To this aim, three types of fibers are considered : aforementioned \ac{SMF}, \ac{LEAF} and Teralight, with variable parameters described in Tab.\ref{tab:fiber_variation}.
\begin{table}[]
	\centering
	
	\begin{tabular}{|l|l|l|l|}
		\hline
		at $\lambda=1550nm$                & SMF  & LEAF & Teralight \\  \hline
		$D$ $(ps\cdot nm^{-1}\cdot km^{-1})$ & 16.8 & 4    & 8         \\ \hline
		$\gamma$ $(W^{-1}\cdot km^{-1})$     & 1.3  & 1.5  & 1.3       \\ \hline
		PMD coefficient $(ps/\sqrt{km})$     & 0.13  & 0.13  & 0.13       \\ \hline
	\end{tabular}
	\caption{Fiber Parameters}
	\label{tab:fiber_variation}
\end{table}

Fig. \ref{fig:mancnlse_fibers}.a and Fig. \ref{fig:mancnlse_fibers}.b show the  $\Delta \mu _{a_{NL}}$ and $\Delta \sigma _{a_{NL}}$ being the differences of $\mu _{a_{NL}}$ and $\sigma _{a_{NL}}$ using either \ac{CNLSE} or Manakov-\ac{PMD} methods, varying the number of plates $N_{p}$. Maximum  differences are observed for \ac{LEAF} then Teralight and finally \ac{SMF}. Figures also indicate that \ac{LEAF} requires more plates to get the same results using both methods (i.e. $N_{p}$ for which $\Delta \mu _{a_{NL}}=0$ ) with respect to Teralight and \ac{SMF}. This is intuitive as the Kerr Length ($L_{Kerr}$) is lower for \ac{LEAF} than for Teralight or \ac{SMF} fibers and thus required shorter $L_{corr.}$ (i.e. larger $N_{p}$) to get an averaged nonlinear effect over polarization states. It is also noticeable that $\Delta \mu _{a_{NL}}$ and $\Delta \sigma _{a_{NL}}$ are higher when using fibers with lower chromatic dispersion coefficient (particularly evident for \ac{SMF} and Teralight since they have the same $\gamma$ coefficient).
\section{Discussion}

In this section, in order to easily integrate the results in the context of optical transmission system domain, we convert the previous results obtained using $a_{NL}$ coefficients in terms of the optimal $Q^2$ factor (noted $Q^2_{opt}$) reachable when being at the optimum launched power in the transmission fibers. Principle of this conversion is detailed in section \ref{sec:gn}. we introduce here two quantities. Firstly $ \delta Q^{2}_{opt} $ defined in Eq. \ref{eq:sigq3} is driven from Eq.\ref{SNRGNAssumptions} and corresponds to the range of variation of the optimum $Q^2$ factor coming from NLI noise and PMD and derived from $a_{NL}$ variation of $\pm 3 \sigma_{aNL}$ around the average value $\mu_{aNL}$.

\begin{equation}
\delta Q^{2}_{opt,CNLSE}(N_p) = \frac{1}{3} \cdot 10 \cdot log_{10}[\frac{\mu_{aNL}(N_p)+3\cdot\sigma_{aNL}(N_p)}{\mu_{aNL}(N_p)-3\cdot\sigma_{aNL}(N_p)}]
\label{eq:sigq3}
\end{equation}

\begin{figure*}[h!]
	\centering
	\begin{tabular}{@{}c@{}}
		\includegraphics[width=.5\linewidth]{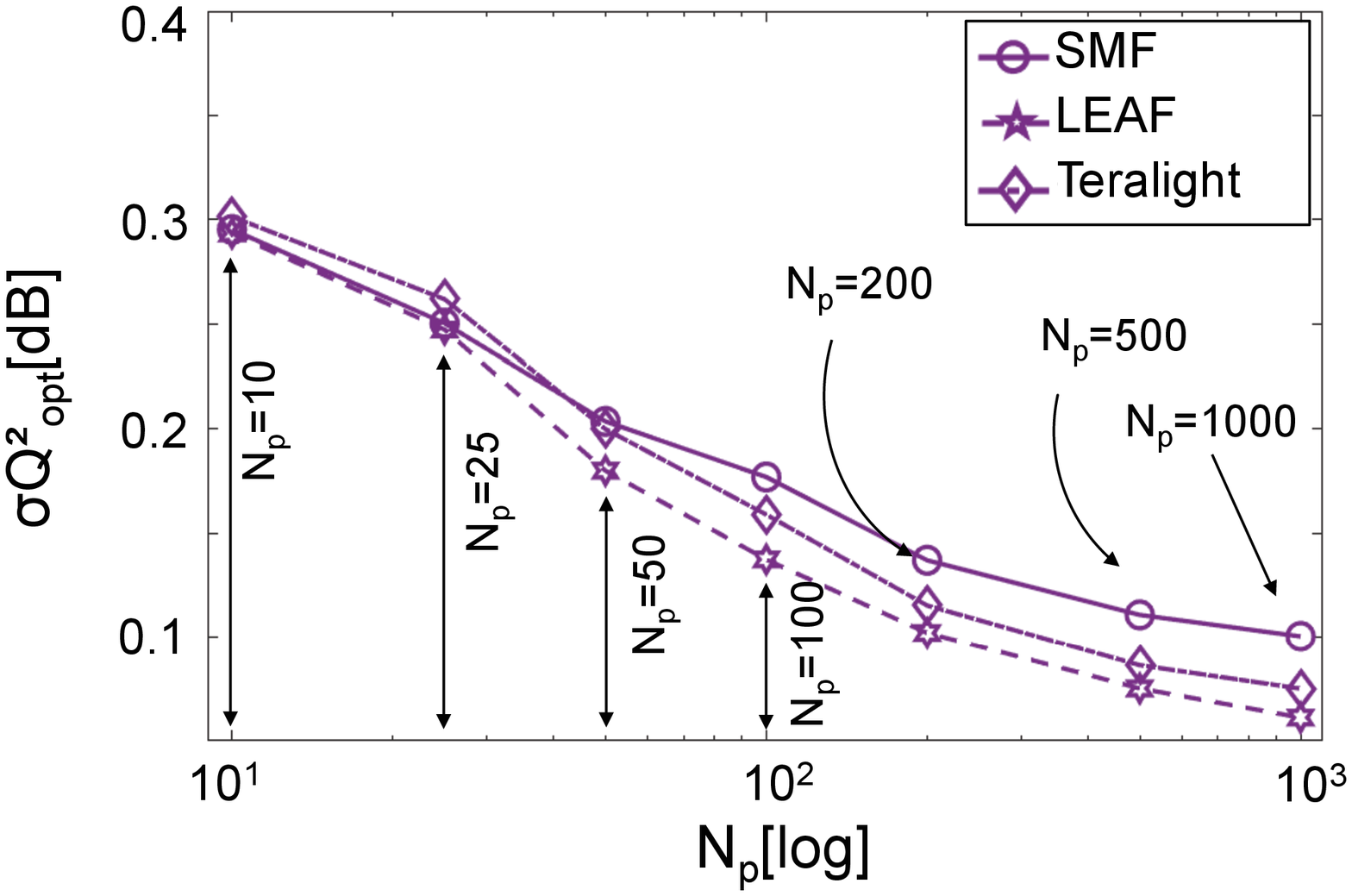}
		\includegraphics[width=.5\linewidth]{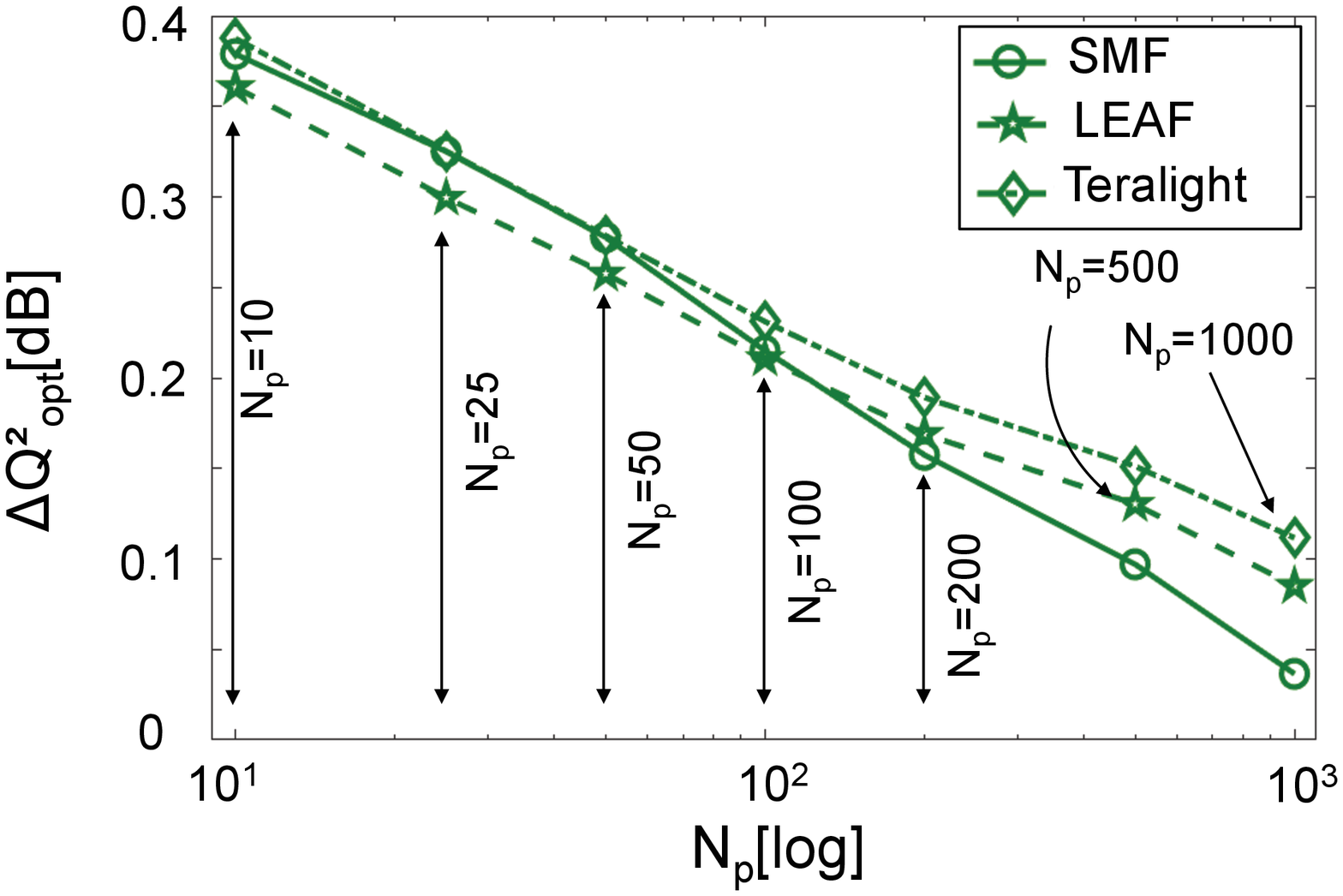}
		\\
		\hspace{1cm}\small (a) \hspace{9cm} \small (b)
	\end{tabular}
		 \caption{(a) $\delta Q^2_{opt,CNLSE}$ and (b) $Q^2_{opt,CNLSE}-Q^2_{opt,Manakov}$ vs. $N_{p}$ using SMF, LEAF and Teralight fiber.}
	\label{fig:q2_cnlse}
\end{figure*} 

Fig. \ref{fig:q2_cnlse}.a shows this $\delta Q^2_{opt}$ from previous \ac{SSFM} simulations using \ac{CNLSE} method depending on the number of birefringent plates. The results demonstrate for instance that considering $N_p=50$ in \ac{CNLSE} method, $\delta Q^2_{opt}$ is about $0.2$ $dB$ for only one span \ac{SMF} transmission. The mentioned $Q^2_{opt}$ variation range for one span \ac{LEAF} and Teralight fiber is about 0.18 and $0.2$ $dB$ respectively considering the same simulation criteria. Increasing the $N_p$ to 1000 plates will reduce $ \delta Q^2_{opt}$ down to less than $0.1$ $dB$ for \ac{SMF} and Teralight, while it is slightly higher for \ac{LEAF}. 

Secondly, we investigate the different absolute values of the $Q^2_{opt}$ considering \ac{CNLSE} method and Manakov-\ac{PMD} method. Fig. \ref{fig:q2_cnlse}.b shows the difference between the $Q^2_{opt}(N_p)$ in \ac{CNLSE} method and  $Q^2_{opt}(50)$ in Manakov-\ac{PMD} method. The figure demonstrates that a difference of $0.3$ $dB$ in $Q^2_{opt}$ between one span \ac{SSFM} simulations using \ac{CNLSE} and Manakov-\ac{PMD} methods for $N_p=50$. 

By taking into account the quantities of Fig. \ref{fig:q2_cnlse}.a and Fig. \ref{fig:q2_cnlse}.b, performing numerical simulations using for instance $N_{p}=50$, the maximum estimation error on the optimal $Q^{2}$ factor can be around $0.5$ $dB$ for just a single-span transmission.

\section{Conclusion}
Quantification and fast estimation of the \ac{QoT} variability experienced by optical fiber systems are essential for the network providers in order to ensure the network reliability by having the system margin knowledge which consequently provides cost reduction. In this work, we quantify the difference between the \ac{CNLSE} and the Manakov-\ac{PMD} methods in terms of \ac{QoT} variability. In order to be compliant with recent fast estimation rules from \ac{GN} modeling, we focus on the estimation of the \ac{NLI} statistics and particularly by assessing the variability of $a_{NL}$ that is a key parameter to deduce the \ac{QoT} for all fiber launched powers and transmission distances. To this aim, we have performed a massive sets of \ac{SSFM} simulation (around 46,800 simulations) to numerically estimate the $a_{NL}$ distribution depending on the stochastic fiber birefringence concatenation using either the \ac{CNLSE} or the Manakov-\ac{PMD} equation. 

Our results quantify the  differences between the $a_{NL}$ distributions obtained by \ac{CNLSE} and Manakov-\ac{PMD} methods as a function of the number of concatenation plates ($N_{p}$) used to emulate the fiber stochastic birefringence. Manakov-\ac{PMD} method is shown to provide constant  results for $N_{p}$ higher than 50 plates. As estimating these distributions is highly time-consuming, it is then advised to perform Manakov-\ac{PMD} based simulation with 50 plates, however the resulting $a_{NL}$ distributions are accurate for the emulation of a fiber with a correlation length ($L_{corr.}$) lower than around 100 meters (or equivalently $N_{p}$>1000). To accurately model fiber with $L_{corr.}$ > 100m, the \ac{CNLSE} method is more accurate, nevertheless an additional time-consumption is needed to emulate the fiber birefringence with higher $N_{p}$. Moreover in this configuration, one can also perform Manakov-\ac{PMD} based simulations and deduce from our results the correction to apply on the averaged $a_{NL}$ ($\mu(a_{NL})$) or its standard deviation ($\sigma(a_{NL})$) to predict more accurate values. We also present an exponential fitting over the results in order to estimate these differences for any $N_p$ and \ac{PMD} values around 0.13 $ps/\sqrt{km}$.    

Finally we have quantified the influence of the fiber type and its \ac{PMD} coefficient on simulation result differences between the $a_{NL}$ distributions obtained by either the \ac{CNLSE} or the Manakov-\ac{PMD} methods while $N_{p}$ is increasing. Results indicates that $a_{NL}$ distributions obtained by \ac{CNLSE} converge toward the ones obtained using Manakov-\ac{PMD} method for lower $N_{p}$ when having a higher \ac{PMD} coefficient, a lower nonlinear coefficient $\gamma$ (or equivalently a higher Kerr nonlinear length) or a higher fiber group velocity dispersion. We have shown that for higher fiber $\gamma$ (i.e. \ac{LEAF}), the difference of the result between the \ac{CNLSE} and the Manakov-\ac{PMD} is larger for any $N_p$ with respect to \ac{SMF} or Teralight fiber. For instance, we have found $\Delta \mu _{a_{NL}}=1.3 \cdot10^{-4}$ $mW^{-2}$ in the \ac{LEAF} considering $N_p=50$, while for the same simulation criteria, $\Delta \mu _{a_{NL}}= 6 \cdot10^{-5}$ $mW^{-2}$ for Teralight and $\Delta \mu _{a_{NL}}= 4 \cdot10^{-5}$ $mW^{-2}$ for \ac{SMF}. In terms of system variability, we have found $\Delta \sigma _{a_{NL}}= 1.3 \cdot10^{-5}$ $mW^{-2}$ for \ac{LEAF}, while the results for Teralight and \ac{SMF} are $7 \cdot10^{-6}$ and $4 \cdot10^{-6}$ $mW^{-2}$ respectively. In the end, in order to discuss the results in a simpler way, we convert the obtained variations of $a_{NL}$ coefficient into optimal $Q^{2}$ factor variation and deduce a potential maximal estimation error of $0.5$ $dB$ while using the less suitable method among \ac{CNLSE} and Manakov-\ac{PMD} and considering a fiber birefringence representation by the concatenation of 50 plates. 

We hope that our investigation will guide system designers to fast and accurately estimate the \ac{QoT} variability of fiber optic transmission system in presence of \ac{PMD} and Kerr nonlinearity.

\section*{Funding}
The research leading to these results was partly funded by DGE (French Government) through the CELTIC+ project SENDATE-TANDEM.

\section*{Acknowledgments}
We want to thank Jehan Procaccia for support on computational servers, Yvan Pointurier for optical networks discussions and Pierre Sillard for giving insights on \ac{PMD}  fiber optics characteristics.

\ifCLASSOPTIONcaptionsoff
  \newpage
\fi




%

\bibliographystyle{IEEEtran}
\bibliography{jocn_v2}

%
%

%




\end{document}